# Systematic Figure of Merit Computation for the Design of Pipeline ADC

L. Barrandon, S. Crand, D. Houzet
IETR, Université de Rennes 1, Avenue du Général Leclerc, 35042 RENNES Cedex, France
ludovic.barrandon@univ-rennes1.fr

## Abstract

*The emerging concept of SoC-AMS leads to research new top-down methodologies to aid systems designers in sizing analog and mixed devices. This work applies this idea to the high-level optimization of pipeline ADC. Considering a given technology, it consists in comparing different configurations according to their imperfections and their architectures without FFT computation or time-consuming simulations. The final selection is based on a figure of merit.*

## 1. Introduction

The evolution of analog and mixed designs and the increasing complexity of such devices lead to develop methodologies using top-down techniques. To illustrate this concept, we have chosen to focus on the performances evaluation of pipeline ADC devices. The aim is to compute a figure of merit (FOM), which consists in integrating various characteristics into a single parameter, to determine the best compromise while matching specifications. Basic constraints consist in saving computation time by avoiding the use of behavioral or electrical level simulations and FFT analysis.

## 2. Figure of merit adjustment

### 2.1 Definition

This FOM is an example adapted here to software defined radio systems where spectral performances (SNDR and SFDR) are dominant [2]. The number of comparators (*Comp*) represents a relative evaluation of the die area and other parameters like the number of residue amplifiers or the evaluation of the power consumption could be considered:

$$FOM = \delta \left[ \alpha \cdot \left( \frac{SNDR}{SNDR_{max}} \right) + \beta \cdot \left( \frac{SFDR}{SFDR_{max}} \right) + \gamma \cdot \left( \frac{Comp_{min}}{Comp} \right) \right]$$

$\alpha = SNDR_{lim} / SNDR_{max}$ $\qquad \gamma = Comp_{min} / Comp_{lim}$
$\beta = SFDR_{lim} / SFDR_{max}$ $\qquad \delta = 1/[3 \cdot (\alpha + \beta + \gamma)]$

$\delta$ is computed to obtain $0 < FOM < 1$

The indices *max* and *min* indicate the optimal values (ideal case) and the indices *lim* represent the desired extrema which the ADC should meet for a given application.

### 2.2 Evaluation of the performances

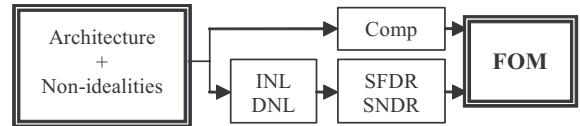

Figure 1. Methodology of evaluation of the performances

Among the different non-idealities of a pipeline ADC stage, only two of them are not corrected by the redundant sign digit (RSD) method: the residue amplification mismatch and the non-linearity of this amplification, whereas offset mismatch on the comparator conversion steps are corrected inside the limit defined as $1/V_{ref} \cdot 2^{Ni+1}$ [2].

**Gain mismatch**: The transfer function $TF_\varepsilon$, taking into account a relative gain mismatch $\varepsilon_{gain}$, is defined from the ideal transfer function *TF* as follows:

$TF_\varepsilon = (1+\varepsilon_{gain}) \cdot TF$

**Non-linearity**: The non–linearity is modeled with a reverse hyperbolic tangent function [3]. The equation used is:

$$TF_{NL} = \frac{1}{\alpha_{NL}} \tanh^{-1}\left( \tanh(\alpha_{NL}) \cdot TF \right)$$

where $\alpha_{NL}$ is a coefficient associated to the amplitude of the non linearity. $1/\alpha_{NL}$ and $\tanh(\alpha_{NL})$ are normalization factors.

**INL calculation:** The INL is the difference between TF and the actual transfer function (figure 2).

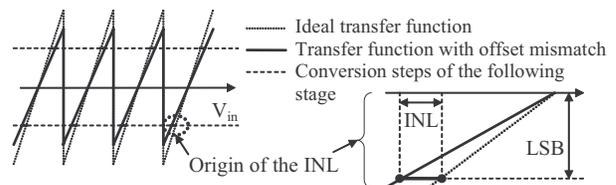

Figure 2. Influence of gain and non-linearity mismatches of a single pipeline stage over the INL

The INL of a non ideal pipeline stage can be written as:

$$INL(i) = \left( (1-\varepsilon_{gain}) \left( \frac{1}{\alpha_{NL}} \tanh^{-1}(\tanh(\alpha_{NL}) x_i) \right) \right) - x_i$$

where $x_i$ is the ideal position of the $i^{th}$ conversion step and $INL(i)$ is the INL corresponding to $x_i$. Then, the global INL of a pipeline ADC is the sum of the contributions of each single stage to this INL.

**SNDR calculation from INL:**

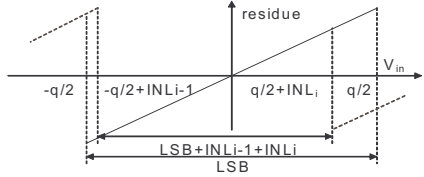

Figure 3. Conversion error for the $i^{th}$ code with and without INL

The SNDR is computed from the mean quadratic error of a non-ideally-quantized sine wave, that is to say adding INL to the ideal conversion steps (figure 3):

$$\overline{\varepsilon_i^2} = \frac{1}{q} \int_{-q/2+INL_{i-1}}^{q/2+INL_i} V_{in}^2 dV_{in} = \frac{1}{3q}\left[\left(\frac{q}{2}+INL_i\right)^3 + \left(\frac{q}{2}-INL_{i-1}\right)^3\right]$$

$$= \frac{q^2}{12} + \frac{q}{4}(INL_i - INL_{i-1}) + \frac{1}{2}(INL_i^2 + INL_{i-1}^2) + \frac{1}{3q}(INL_i^3 - INL_{i-1}^3)$$

where $q^2/12$ is equal to the well-known conversion noise for an ideal ADC. Then, the SNDR of an N-bit ADC becomes:

$$SNDR = 20.\log\left(\frac{V_{ref}^2/2}{\overline{\varepsilon^2}/2^N}\right) \quad \text{with} \quad \overline{\varepsilon^2} = \sum_i \overline{\varepsilon_i^2}$$

**SFDR calculation from INL**: In [4], the levels of the harmonics of an ideally-quantized sine-wave are computed from Fourier series expansion:

$$a_k = \frac{2}{\pi k} \sum_{i=1}^{2^n} y_i \left[\sin(k.\cos^{-1}(x_i)) - \sin(k.\cos^{-1}(x_{i+1}))\right]$$

where $x_i$ are the thresholds and $y_i=(x_i+x_{i+1})/2$. The originality of this work is to add the DNL to the ideal quantization steps. The SFDR is the highest harmonic level given by:

$$HD_k \equiv 20.\log(|a_k/a_1|) \quad dB_c$$

## 3. Results and interpretations

In this study, pipeline ADC use the RSD method to enable digital correction [2]. The use of 1.5-bit stages is allowed, except for the first stage (where digital correction is inefficient as the output varies from $-FS/2$ to $FS/2$ with a full-scale input) and for the last stage which is a simple flash ADC. Table 1 presents the extrema obtained with the FOM program for 10-bit ADC (1596 possibilities) with:

$\varepsilon_{gain} = -1.5\%$ ; $\alpha_{NL} = 0.2$
$Comp_{lim} = 60 = 12\%$ of $Comp_{max}$
$SFDR_{lim} = 75dB = 88\%$ of $SFDR_{max}$
$SNDR_{lim} = 56dB = 90\%$ of $SNDR_{max}$.

| Comp | SNR(dB) | ENOB(Bits) | SFDR(dB) | FOM |
|---|---|---|---|---|
| 2/2/2/7 | 136 | 49,7 | 7,97 | 81,7 | **0,77** |
| 2/9 | **514** | 53,0 | 8,51 | 84,5 | 0,79 |
| 2/2/2/2/2/2/2/2 | 27 | **49,4** | **7,91** | 79,7 | 0,85 |
| 9/2 | **514** | **62,0** | **10,00** | 84,6 | 0,85 |
| 2/1.5/2/2/2/2/2/1.5/2 | 25 | 51,0 | 8,17 | **79,1** | 0,87 |
| 2/1.5/1.5/1.5/1.5/1.5/1.5/1.5/2 | **20** | 52,9 | 8,49 | 84,9 | 0,94 |
| 4/1.5/1.5/2/1.5/1.5/2 | 29 | 60,0 | 9,68 | **86,7** | 0,95 |
| 3/1.5/1.5/1.5/1.5/1.5/1.5/2 | 22 | 57,5 | 9,26 | 85,9 | **0,96** |

$\alpha= 0.909$; $\beta= 0.885$; $\gamma= 0.333$; $\delta= 0.470$

Table 1. Extrema obtained with the FOM program for 10-bit pipeline ADC

These results are representative of the following tendencies. 9/2 configuration is the closest to the flash architecture. Its performances are the best in terms of SNDR and SFDR but its number of comparators is also the largest. The architecture only composed of 2-bit stages has the worst performances. Nevertheless, thanks to its reduced number of comparators, this solution does not lead to the lowest FOM. The best architecture, according to the FOM, is an ADC composed of an N-bit front-end stage, with N an integer scaled according to $\varepsilon_{gain}$ et $\alpha_{NL}$ (i.e. N=3 for this case), followed by 1.5-bit stages and finished by a 2-bit stage. This analysis is confirmed for other resolutions and other values of $\varepsilon_{gain}$ and $\alpha_{NL}$.

## 4. Conclusion and perspectives

This work presents an efficient way to exhaustively arrange pipeline architectures according to their estimated performances. This method is applicable to other analog or mixed devices with a SoC-AMS top-down design perspective [1]. It will be used to facilitate the exploration of the design space of the mixed front-end for software defined radio applications [5].

The main aspect which remains to develop is the estimation of $\varepsilon_{gain}$ and $\alpha_{NL}$ parameters according to actual technologies [3].

## References

[1] M. Vogels, G. Gielen "Figure of merit based selection of A/D converters", DATE 2003, CAD for Analogue Design, Design Methodologies and Physical Design p.1190
[2] L. Sumanen, "Pipeline analog-to-digital converters for wideband wireless communications", Helsinki University of Technology Electronic Circuit Design, Report 35, 2002
[3] P. Nuzzo, "Architectural Space Characterization of a Transconductance Amplifier", Università di Pisa, 2003
[4] Hui Pan, "A 3.3-V, 12-bit, 50-MS/s A/D Converter in 0.6-μm CMOS", University of California, Los Angeles, PhD thesis, 1999
[5] L. Barrandon, S. Crand, D. Houzet, "Behavioral Modeling and Simulation of Mixed Signal Front-End for Software Defined Radio Terminals", ISIE 2004, Ajaccio – France